\documentclass[sigconf]{acmart}

\usepackage{amsmath}
\usepackage{amsthm}
\usepackage{subfigure}
\usepackage{graphicx}
\usepackage{enumitem}
\usepackage{multirow}
\usepackage{makecell}
\usepackage{graphicx}
\usepackage{subfloat}
\usepackage{amsfonts}
\usepackage{color}
\usepackage{caption}
\usepackage{subcaption}
\usepackage{enumitem}
\usepackage{xspace}
\usepackage{xcolor} 
\usepackage{mdframed}
\usepackage[ruled]{algorithm2e}

\AtBeginDocument{%
  }

\setcopyright{acmlicensed}
\copyrightyear{2026}
\acmYear{2026}
\setcopyright{cc}
\setcctype{by}
\acmConference[KDD '26]{Proceedings of the 32nd ACM SIGKDD Conference on Knowledge Discovery and Data Mining V.1}{August 09--13, 2026}{Jeju Island, Republic of Korea}
\acmBooktitle{Proceedings of the 32nd ACM SIGKDD Conference on Knowledge Discovery and Data Mining V.1 (KDD '26), August 09--13, 2026, Jeju Island, Republic of Korea}
\acmPrice{}
\acmDOI{10.1145/3770854.3783917}
\acmISBN{979-8-4007-2258-5/2026/08}
\begin{document}


\title{Optimizing Generative Ranking Relevance via Reinforcement Learning in Xiaohongshu Search}


% \affiliation{%

\author{Ziyang Zeng}
\authornote{This work was conducted during Ziyang's internship at Xiaohongshu.}
\authornote{Both authors contributed equally to this work.}
\orcid{0009-0000-8773-7301}
\affiliation{%
  \institution{Beijing University of Posts and Telecommunications}
  \city{Beijing}
  \country{China}
}
\email{ziyang1060@bupt.edu.cn}

\author{Heming Jing}
\authornotemark[2]
\orcid{0000-0001-9216-7032}
\email{jingheming@xiaohongshu.com}
\author{Jindong Chen}
\orcid{0009-0003-2977-398X}
\email{chenjindong@xiaohongshu.com}
\affiliation{%
  \institution{Xiaohongshu Inc.}
  \city{Beijing}
  \country{China}
}

\author{Xiangli Li}
\orcid{0000-0003-0126-932X}
\email{lixiangli@xiaohongshu.com}
\author{Hongyu Liu}
\orcid{0009-0001-8009-5112}
\email{liuhongyu1@xiaohongshu.com}
\affiliation{%
  \institution{Xiaohongshu Inc.}
  \city{Beijing}
  \country{China}
}

\author{Yixuan He}
\orcid{0009-0003-0566-0622}
\email{heyixuan@xiaohongshu.com}
\author{Zhengyu Li}
\orcid{0000-0003-2836-276X}
\email{luminous@xiaohongshu.com}
\affiliation{%
  \institution{Xiaohongshu Inc.}
  \city{Beijing}
  \country{China}
}

\author{Yige Sun}
\orcid{0009-0004-3025-2142}
\email{sunyige@xiaohongshu.com}
\author{Zheyong Xie}
\orcid{0009-0009-7453-5781}
\email{xiezheyong@mail.ustc.edu.cn}
\affiliation{%
  \institution{Xiaohongshu Inc.}
  \city{Beijing}
  \country{China}
}

\author{Yuqing Yang}
\authornote{Corresponding authors.}
\orcid{0000-0001-5333-1346}
\affiliation{%
  \institution{Beijing University of Posts and Telecommunications}
  \city{Beijing}
  \country{China}
}
\email{yangyuqing@bupt.edu.cn}

\author{Shaosheng Cao}
\orcid{0000-0002-3795-8824}
\email{shelsoncao@gmail.com}
\author{Jun Fan}
\orcid{0009-0000-2127-0702}
\authornotemark[3]
\email{shenhai1@xiaohongshu.com}
\affiliation{%
  \institution{Xiaohongshu Inc.}
  \city{Beijing}
  \country{China}
}

\author{Yi Wu}
\orcid{0009-0007-8838-2785}
\affiliation{%
  \institution{Xiaohongshu Inc.}
  \city{Beijing}
  \country{China}
}
\email{luyun2@xiaohongshu.com}

\author{Yao Hu}
\orcid{0009-0006-1274-7111}
\affiliation{%
  \institution{Xiaohongshu Inc.}
  \city{Beijing}
  \country{China}
}
\email{xiahou@xiaohongshu.com}

% \renewcommand{\shortauthors}
\renewcommand{\shortauthors}{Ziyang Zeng et al.}

\begin{abstract}
Ranking relevance is a fundamental task in search engines, aiming to identify the items most relevant to a given user query. Traditional relevance models typically produce scalar scores or directly predict relevance labels, limiting both interpretability and the modeling of complex relevance signals. Inspired by recent advances in Chain-of-Thought (CoT) reasoning for complex tasks, we investigate whether explicit reasoning can enhance both interpretability and performance in relevance modeling. However, existing reasoning-based Generative Relevance Models (GRMs) primarily rely on supervised fine-tuning on large amounts of human-annotated or synthetic CoT data, which often leads to limited generalization. Moreover, domain-agnostic, free-form reasoning tends to be overly generic and insufficiently grounded, limiting its potential to handle the diverse and ambiguous cases prevalent in open-domain search. In this work, we formulate relevance modeling in Xiaohongshu search as a reasoning task and introduce a Reinforcement Learning (RL)-based training framework to enhance the grounded reasoning capabilities of GRMs. Specifically, we incorporate practical business-specific relevance criteria into the multi-step reasoning prompt design and propose Stepwise Advantage Masking (SAM), a lightweight process-supervision strategy which facilitates effective learning of these criteria through improved credit assignment. To enable industrial deployment, we further distill the large-scale RL-tuned model to a lightweight version suitable for real-world search systems. Extensive offline evaluations and online A/B tests demonstrate that our approach consistently delivers significant improvements across key relevance and business metrics, validating its effectiveness, robustness, and practicality for large-scale industrial search systems.
% Extensive experiments on industrial datasets, along with online A/B tests, demonstrate the effectiveness of our approach.

\end{abstract}



%   \includegraphics[width=\textwidth]{case.pdf}


\maketitle

\section{Introduction}
Search engines~\cite{croft2010search} such as Google and Xiaohongshu\footnote{\url{https://www.xiaohongshu.com/explore}} have become indispensable gateways to online information, serving hundreds of millions of user queries every day.
A key enabler of their effectiveness is the accurate assessment of ranking relevance~\cite{YahooSearch}, which fundamentally underpins improvements in user experience and satisfaction.
Traditional relevance modeling approaches predominantly adopt scalar discriminative models~\cite{bertrank,yates-etal-2021-pretrained,BaiduSearch}, producing a single probability that indicates whether a query-item pair is relevant.
Despite their success, these models operate as ``black boxes’’, offering limited interpretability and making it difficult to diagnose relevance errors. 
Moreover, their limited knowledge capacity restricts their ability to understand the complex and nuanced semantics underlying relevance.

The rapid progress of Large Language Models (LLMs) in industrial applications~\cite{Darec,GQS} has sparked the development of Generative Relevance Models (GRMs)~\cite{zhu2024largelanguagemodelsinformation}.
By keeping LLMs' output head and leveraging their generative capabilities, GRMs enable more flexible and interpretable relevance assessments.
Existing GRMs fall into two main categories:
(1) \emph{Vanilla GRMs}, which generate a single token (e.g., ``Yes'' or ``No'') to indicate relevance~\cite{LLaMARanking}. 
However, such single-token decisions impose a strict token budget that limits semantic elaboration, leading to suboptimal performance in complex relevance scenarios.
(2) \emph{Reasoning-based GRMs}, which incorporate Chain-of-Thought (CoT) reasoning~\cite{cot} before producing relevance judgments~\cite{niu2024judgerankleveraginglargelanguage,weller2025rank1testtimecomputereranking}, and have demonstrated promising results in vertical domains such as e-commerce search~\cite{Ecommerce1,Ecommerce2}.
Despite their success, these reasoning-based models predominantly rely on Supervised Fine-Tuning (SFT) using costly human-annotated or synthetic CoT data distilled from advanced LLMs~\cite{Ecommerce1,Ecommerce2}, which are data-intensive and exhibit limited generalization~\cite{chu2025sftmemorizesrlgeneralizes}.
Moreover, relevance modeling in open-domain search remains particularly challenging due to the diversity of user queries and the prevalence of ambiguous search cases for which domain-agnostic, free-form reasoning fails to yield clear relevance judgments.
Additionally, unlike vertical search domains enriched with structured signals (e.g., product brands and categories), open-domain search must operate over unstructured content, which significantly increases the difficulty of CoT reasoning and undermines its consistency and reliability in the absence of domain-specific guidance~\cite{10.1145/3701551.3703583}.

\begin{figure}[!t]
    \centering
    \includegraphics[width=\linewidth]{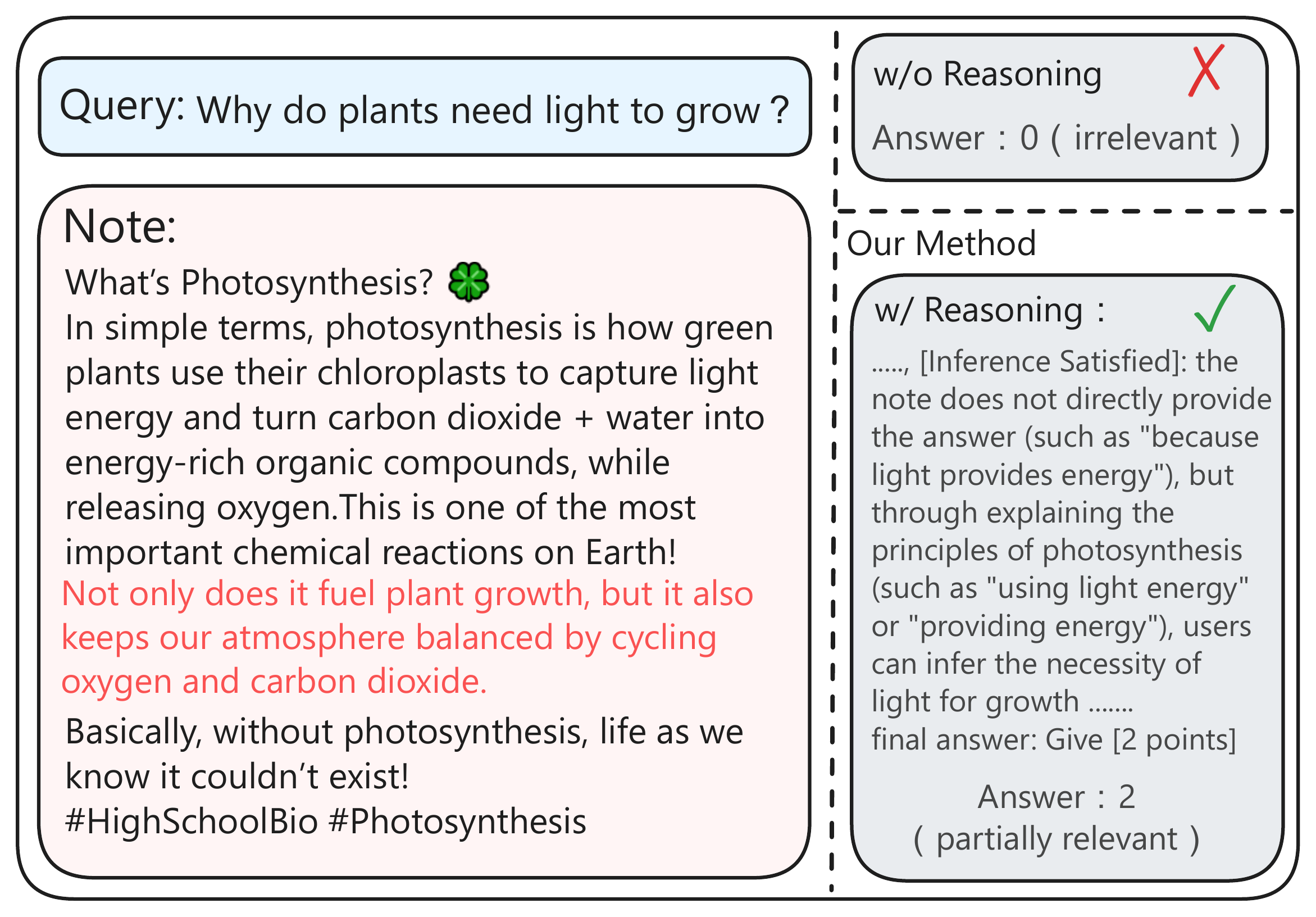}
    \caption{An example illustrating that explicit reasoning enhances both interpretability and effectiveness of relevance assessment. For the query ``Why do plants need light to grow?'', a reasoning-based model leverages photosynthesis-related content to recognize partial relevance, whereas a model without reasoning fails to identify this connection.
    }
    \label{fig:case}
\end{figure}

\begin{figure*}[!h]
    \centering
    \includegraphics[width=\textwidth]{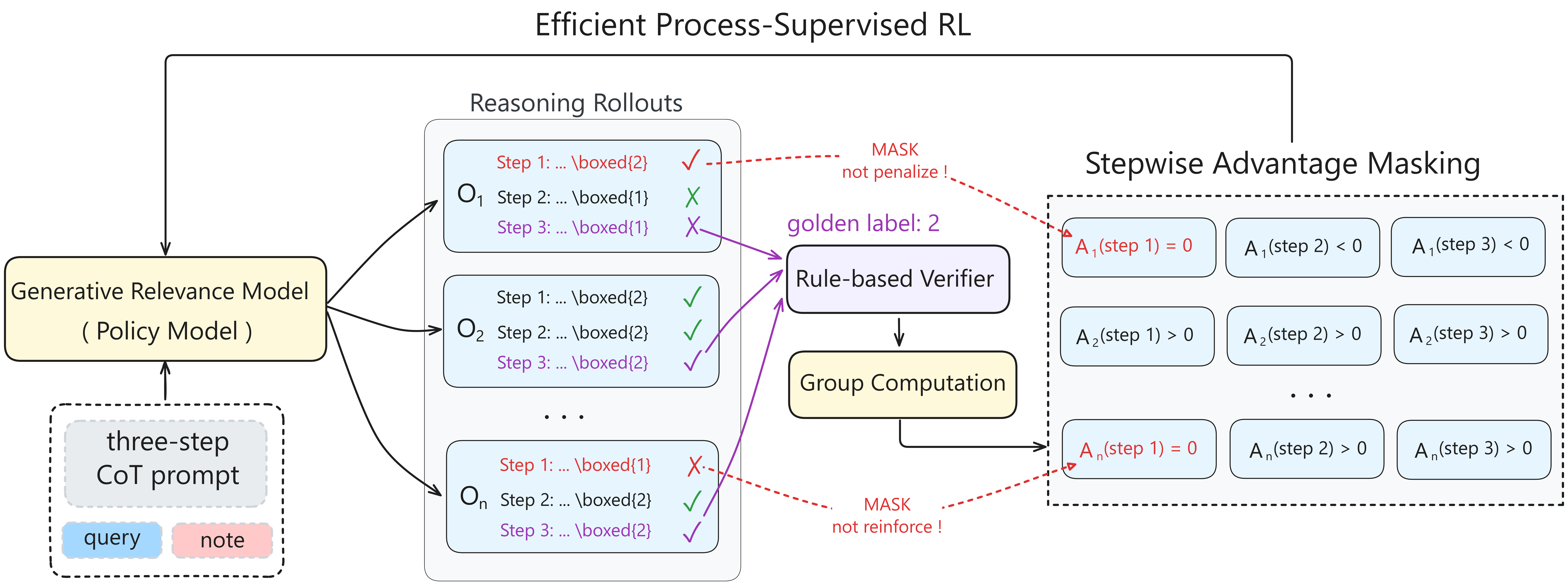}
    \caption{
    Illustration of the Stepwise Advantage Masking (SAM) strategy for process-supervised reinforcement learning. In the three-step relevance reasoning, the model produces intermediate scores (denoted by \texttt{\textbackslash boxed\{\}}) at each step, which are validated against the ground-truth label using a rule-based verifier (i.e., exact matching). We define correctness indicators $(c_1, c_2, c_3)$, where $c_i = \texttt{True}$ if step $i$ yields a correct intermediate prediction. 
    SAM leverages these indicators to construct a stepwise advantage mask: if the final answer is correct, only correct steps are reinforced; if the final answer is incorrect, only the erroneous steps are penalized.
    This selective credit assignment prevents spurious reward propagation and facilitates efficient step-level optimization of generative relevance models.
    }
    \label{fig:framework}
\end{figure*}

In this work, we formulate relevance modeling in Xiaohongshu search as a reasoning task and introduce a new Reinforcement Learning (RL)-based training paradigm that enhances the grounded reasoning capabilities of GRMs, yielding substantial improvements in both interpretability and performance (see Figure~\ref{fig:case}).
To overcome the limitations of domain-agnostic reasoning in open-domain search, we incorporate domain-specific relevance criteria—accumulated through years of Xiaohongshu search system development—directly into the multi-step reasoning prompt design.
By injecting these well-established decision rules, particularly those tailored for challenging or ambiguous cases, we guide the model’s reasoning process with critical prior knowledge, thereby aligning its relevance judgments more closely with real-world user expectations.
To ensure that the model effectively internalizes these criteria, we further propose Stepwise Advantage Masking (SAM), a lightweight process supervision strategy that improves credit assignment in RL\footnote{Credit assignment refers to the problem of accurately attributing success or failure to individual actions within a sequential decision process, particularly when rewards are sparse or delayed~\cite{712192}.} (see Figure~\ref{fig:framework}).
SAM employs a rule-based verifier over intermediate reasoning steps to provide targeted supervision and distribute rewards more accurately along the reasoning trajectory.
This efficient approach eliminates the need for expensive full-trajectory human annotations~\cite{lightman2024lets} or  highly computational Monte Carlo-based online value estimation~\cite{kazemnejad2025vineppo}, making it a practical process-supervised RL solution for industrial systems.
 
To enable practical deployment, we adopt a distillation-based approach to adapt our large-scale RL-tuned model for Xiaohongshu’s search production environment. 
Extensive offline experiments on industrial datasets, along with online A/B
tests, demonstrate that our approach consistently outperforms strong baselines and offers a practical design paradigm for RL-based relevance modeling.

Our key contributions are summarized as follows:
\begin{itemize}
    \item We formulate relevance modeling as a reasoning task and introduce a novel RL-based training paradigm that achieves stronger generalization than the data-intensive SFT baseline.
    \item We incorporate domain-specific relevance criteria, derived from Xiaohongshu’s industrial search practices, into the reasoning prompt design, providing essential inductive bias in ambiguous search scenarios.
    \item We propose Stepwise Advantage Masking (SAM), a lightweight process supervision strategy that enables step-level credit assignment in RL without costly human annotations or online value estimation.
    \item We validate our approach on Xiaohongshu search through extensive offline experiments and online tests, achieving substantial gains over baselines.
\end{itemize}

\section{Related Work}
\subsection{Generative Relevance Models}

The rapid advancement of LLMs has shifted relevance modeling from traditional discriminative classification~\cite{bertrank,yates-etal-2021-pretrained,rankt5} to generative paradigms (i.e., GRMs)~\cite{zhuang-etal-2024-beyond,sun-etal-2023-chatgpt}. 
GRMs exploit the text generation capabilities of LLMs to assess query-item relevance in a more flexible and interpretable manner.
Early GRMs follow a pointwise scoring strategy, deriving relevance from the likelihood of specific token outputs (e.g., ``Yes'' or ``No'')~\cite{liang2023holistic,zhuang-etal-2024-beyond}. 
Recent progress introduces ranking-oriented prompting strategies that support pairwise~\cite{qin-etal-2024-large} and listwise~\cite{sun-etal-2023-chatgpt} comparisons, benefiting from LLMs' instruction-following abilities. 
To further enhance relevance assessment, CoT prompting has been incorporated into GRMs~\cite{niu2024judgerankleveraginglargelanguage,weller2025rank1testtimecomputereranking}, enabling multi-step reasoning akin to human judgment.
However, existing reasoning-based approaches either depend on SFT with costly collected reasoning paths~\cite{Ecommerce2,Ecommerce1,weller2025rank1testtimecomputereranking}, which may generalize poorly to long-tail queries, or adopt outcome-based RL~\cite{zhuang2025rankr1enhancingreasoningllmbased,zhang-etal-2025-rearank} without fine-grained supervision of intermediate reasoning steps, leading to suboptimal reasoning chains.
In contrast, we introduce a process-supervised RL framework that performs step-level credit assignment, yielding more grounded reasoning trajectories and significantly improving robustness in ambiguous search scenarios.

\subsection{Criteria-augmented Relevance Modeling}
Recent advances in relevance modeling highlight the importance of incorporating explicit criteria to guide ranking decisions. 
Early relevance criteria are often formulated as information retrieval axioms, which define desirable ranking properties through formal mathematical constraints to guide the design of ranking models. 
% Early relevance criteria are often formulated as information retrieval axioms, mathematical constraints that impose desirable ranking behaviors.
Prior work demonstrates the benefits of augmenting neural rankers with axiomatic regularization during fine-tuning~\cite{10.1145/3331184.3331296}, while ARES~\cite{10.1145/3477495.3531943} further embeds these constraints into the pre-training phase to enhance ad hoc retrieval performance. 
Other efforts leverage axiomatic perturbations to synthesize training data that improves neural ranking~\cite{10.1145/3409256.3409828}.
With the emergence of LLM-based ranking, the notion of relevance criteria has evolved from rigid symbolic axioms to natural-language formulations. 
MCRanker~\cite{10.1145/3701551.3703583} introduces automatically generated multi-perspective criteria to guide zero-shot pointwise LLM rankers during inference. 
In the medical domain, \citet{zeng2025explainabledoctorrecommendationlarge} constructs disease-specific synthetic criteria to enhance the consistency and interpretability of LLM-based doctor ranking.
However, automatically induced criteria often suffer from semantic drift, limited domain reliability, and insufficient alignment with real-world user intent. 
Unlike prior work, our approach leverages high-precision, expert-curated relevance criteria accumulated through years of industrial development at Xiaohongshu search. 
These criteria inherently reflect authentic intent patterns and ambiguity-resolution strategies observed in real search scenarios, providing stronger practical grounding and reliability than synthetic alternatives.
To operationalize such criteria within LLMs, we further employ process-supervised RL training, enabling the model not only to memorize the rules but also to apply them coherently during step-by-step reasoning. 
% This design leads to improved generalizability, robustness under distribution shifts, and controllable relevance behavior in open-domain search.

\section{Preliminaries}
Relevance modeling in Xiaohongshu search is defined as a multi-class classification task, whose objective is to assign a discrete relevance label \( y \in \{-1, 0, 1, 2, 3\} \) to each query-note pair \((q, n)\). 
Each label corresponds to a distinct level of semantic relevance, ranging from strongly irrelevant (\(-1\)), irrelevant (\(0\)), weakly relevant (\(1\)), partially relevant (\(2\)), to perfectly relevant (\(3\)).
The annotation scheme is derived from human judgments made by trained evaluators, following detailed guidelines that account for search intent alignment, topical consistency, expected user engagement, and other relevance-related dimensions.

\section{Methodology}
This section outlines our RL-based training paradigm.
Section~\ref{sec:reltask} reformulates relevance modeling from a generative perspective, enabling the model to jointly produce reasoning traces and final assessments.
Section~\ref{sec:prompt} introduces our criteria-augmented prompt design, while Section~\ref{sec:distill} presents a distillation-driven cold-start initialization that stabilizes subsequent RL training.
Finally, Section~\ref{sec:rl} details our process-supervised RL framework and its Stepwise Advantage Masking (SAM) mechanism, which further improves reasoning fidelity and quality.

\subsection{Relevance Modeling as Reasoning}
\label{sec:reltask}
In this work, we formulate relevance modeling in Xiaohongshu search as a reasoning task rather than a direct prediction problem.
To achieve deeper semantic understanding and provide interpretable decision paths, we introduce a generative relevance model \(\pi_\theta\)  that first performs structured reasoning to produce a reasoning trace \(\mathbf{o} = (o_1, o_2, \ldots, o_T)\), and then derives the final relevance label \(\hat{l}\) based on this reasoning process.
Given a dataset 
\(\mathcal{D} = \{(p^{(i)}, q^{(i)}, n^{(i)}, l^{(i)})\}_{i=1}^N,\)
where \(p\) is the task prompt (instruction), \(q\) is the query, \(n\) is the note, and \(l\) is the golden relevance label, 
the reasoning trace $\mathbf{o}$ is generated in an autoregressive manner as:
\begin{equation}
\pi_\theta(\mathbf{o} \mid p, q, n) 
= \prod_{t=1}^T \pi_\theta(o_t \mid p, q, n, o_{<t}).
\end{equation}
Here, \(\pi_\theta(o_t \mid p, q, n, o_{<t})\) denotes the probability of generating the \(t\)-th token in the reasoning trace,
conditioned on the input \((p, q, n)\) and all previously generated tokens \(o_{<t}\).
The output \(\mathbf{o}\) includes both the reasoning process and the predicted label 
\(\hat{l} \in \mathbf{o}\).
Our objective is to maximize the probability that the generated label matches the ground-truth label:
\begin{equation}
\max_{\pi_\theta} \mathbb{E}_{(p, q, n, l) \sim \mathcal{D}, 
\mathbf{o} \sim \pi_\theta( \cdot \mid p, q, n)}
\left[ \mathbb{I}(f(\mathbf{o}) = l)\right],
\end{equation}
where \(\mathbb{I}(\cdot)\) is the indicator function, and $f(\mathbf{o})$ extracts the final predicted label $\hat{l}$ from the generated reasoning trace $\mathbf{o}$.

\subsection{Criteria-augmented Prompt Design}
\begin{figure}[t]
    \centering
    \includegraphics[width=\columnwidth]{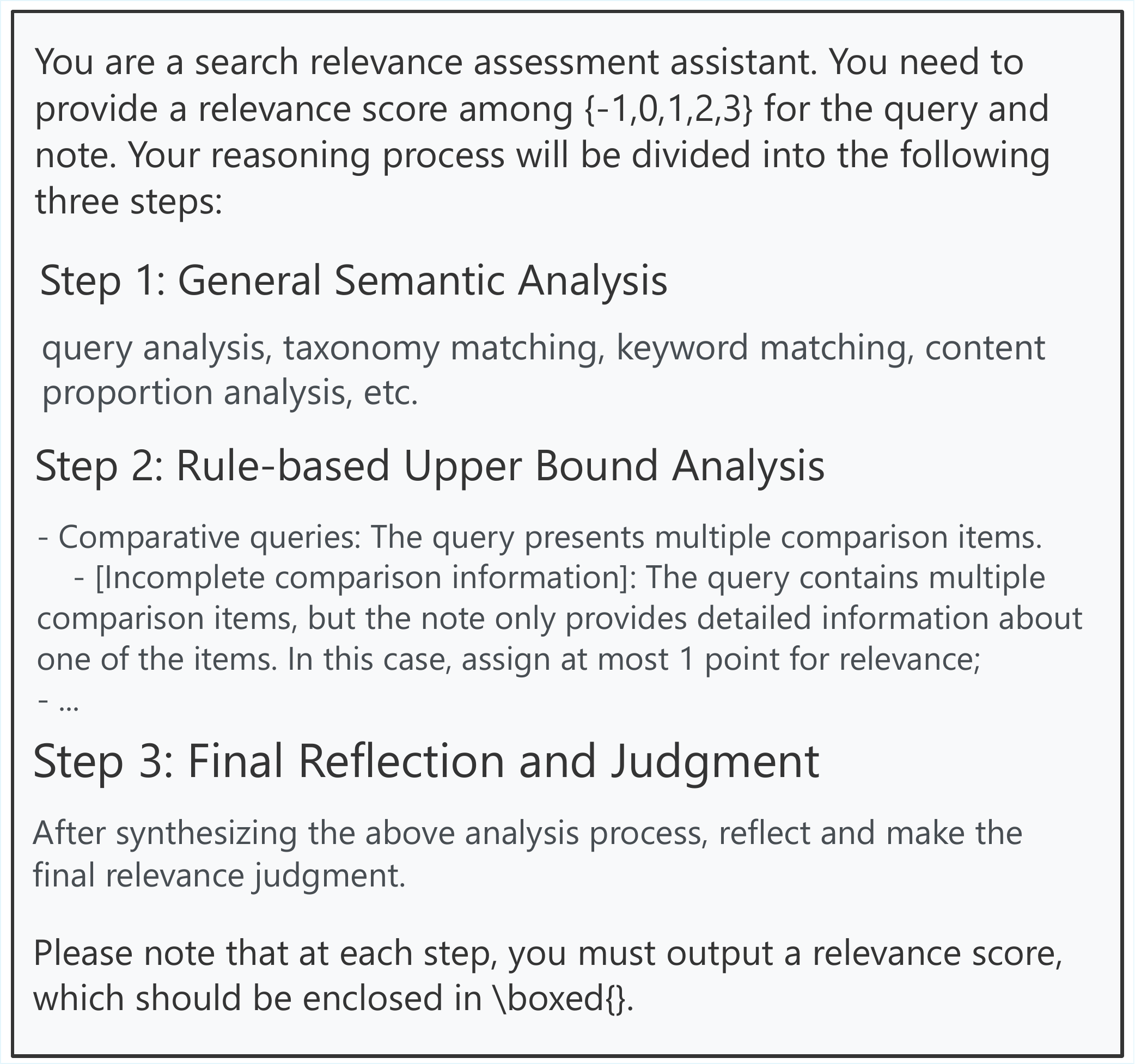}
    \caption{Prototype of the criteria-augmented prompt.}
    \label{fig:prompt}
\end{figure}

\label{sec:prompt}
While recent advances in RL with verified rewards have demonstrated strong capabilities in mathematical reasoning~\cite{r1}, extending this paradigm to search relevance assessment remains fundamentally challenging.
Mathematical and coding tasks are inherently \emph{objective}, governed by universal axioms and formal logic.
Since LLMs already acquire the foundational knowledge for such tasks during pretraining, RL can further enhance performance by promoting unconstrained reasoning grounded in this prior~\cite{yue2025does}.
In contrast, relevance assessment is intrinsically \emph{subjective} and \emph{domain-specific}. 
The relevance criteria are tightly coupled with the specific business context and user ecosystem of platforms like Xiaohongshu, especially in ambiguous search cases.
Without specific guidance, a model cannot reliably infer such nuanced criteria through unguided exploration, often leading to suboptimal performance.
To bridge this gap, we leverage Xiaohongshu’s relevance criteria, a structured body of operational knowledge developed over years of search optimization.
These criteria include general relevance principles defined by human experts, as well as specialized rules for handling edge cases, which together serve as pseudo-axioms to guide the model's reasoning process.
Building on these, we design a structured prompt (see Figure~\ref{fig:prompt}) that internally decomposes the relevance assessment into three interpretable reasoning steps:

\begin{enumerate}
\item \textbf{General Semantic Analysis:}
Following general relevance principles, the model conducts a high-level semantic comparison between the query $q$ and the note $n$, including query analysis, taxonomy and keyword matching, content proportion analysis, and more. This step produces an initial relevance score from a broad perspective.

\item \textbf{Rule-based Upper Bound Analysis:} 
According to the specialized rules for nuanced cases, the model determines the upper bound of the relevance score by applying the rules that match the current scenario. If multiple rules are triggered, the model further tightens the constraint by selecting the smallest resulting upper bound.

\item \textbf{Final Reflection and Judgment:}  
The model reflects on the outputs from the previous steps, integrating both the initial relevance estimate and the upper-bound constraint, to derive the final relevance label $\hat{l}$.
\end{enumerate}

Notably, both Step~1 and Step~2 are capable of generating independent relevance estimates. However, these two estimates serve different purposes: the first provides an initial relevance assessment, while the second supplies an upper bound on the achievable relevance. In practice, the final relevance judgment is derived by combining the general assessment with the upper bound constraint obtained in the second step.
See Appendix~\ref{apx:effect_rel_cri} for empirical validation of the proposed criteria-augmented prompt.

\subsection{Distillation-based SFT for Warm-up}
\label{sec:distill}
Preliminary experiments reveal that directly applying RL from a cold start leads to unstable reasoning behaviors: the model often fails to follow the multi-step instructions specified in our prompt.
To mitigate this cold-start issue, we bootstrap the reasoning ability of GRMs through a SFT stage using synthesized reasoning traces. This warm-up step serves as actor initialization for the subsequent RL phase.
Given each example \((p^{(j)}, q^{(j)}, n^{(j)}, l^{(j)}) \in \mathcal{D}_{\text{distill}}\), we prompt advanced LLMs to generate a structured reasoning trace \(\mathbf{\hat{o}}^{(j)}\) using the multi-step prompt introduced in Section~\ref{sec:prompt}. This prompt enforces a three-step reasoning format, with intermediate boxed scores \verb|\boxed{}| at the end of each step.
To ensure high-quality imitation learning, we employ rejection sampling: a reasoning trace is accepted only if the final predicted relevance label \(\hat{l}\) (extracted from the third step of \(\mathbf{\hat{o}}\)) matches the ground-truth label \(l\).
After filtering, the final distillation dataset is defined as $\mathcal{D'}_{\text{distill}} = \left\{ (p^{(j)}, q^{(j)}, n^{(j)}, \mathbf{\hat{o}}^{(j)}) \right\}_{j=1}^M$.
We train the model by minimizing the Negative Log-Likelihood (NLL) loss:
\begin{equation}
\mathcal{L}_{\text{distill}}(\theta) = - \sum_{(p,q,n,\mathbf{\hat{o}}) \in \mathcal{D'}_{\text{distill}}} \sum_{t=1}^T \log \pi_{\theta} (\hat{o}_t \mid p, q, n, \hat{o}_{<t}).
\end{equation}
This objective guides the model to accurately reproduce the reference reasoning traces under a teacher-forcing regime.

\subsection{Process-supervised RL Training}
\label{sec:rl}
To overcome the data-intensive nature and limited generalization of SFT, we introduce RL to enhance the model’s grounded reasoning capabilities. Specifically, we treat the GRM \(\pi_{\theta}(\cdot \mid p, q, n)\) as a policy model, and optimize the following objective:
\begin{equation}
\label{eq:rl}
\max_{\pi_{\theta}} \, \mathbb{E}_{(p, q, n, l) \sim \mathcal{D}, \mathbf{o} \sim \pi_{\theta}(\cdot \mid p, q, n)}
\big[ \mathcal{R}(\mathbf{o},l)  - \beta\;\mathbb{D}_{\mathrm{KL}}(\pi_{\theta} \| \pi_{\mathrm{ref}}) \big],
\end{equation}
where \(\pi_{\mathrm{ref}}\) is the reference model (i.e., the model before RL fine-tuning), \(\mathcal{R}(\mathbf{o},l)\) is the reward function, \(\mathbb{D}_{\mathrm{KL}}\) is the KL divergence, and \(\beta\) is a hyperparameter balancing the task-specific objective with the KL regularization.

\subsubsection{\textbf{Reward Design}}
Rule-based reward mechanisms are widely used in RL for verifiable tasks due to their unbiased, efficient, and intuitive characteristics. 
In our training, we adopt a simple correctness-based reward, which directly reflects the model's accuracy in assessing relevance. Formally, our reward is defined as follows:
\begin{equation}
\mathcal{R}(\mathbf{o},l) =
\begin{cases}
1, & \text{if } f(\mathbf{o}) = \hat{l}\; \text{and }\;\hat{l} = l, \\
0, & \text{otherwise}.
\end{cases}
\label{eq:reward}
\end{equation}

\subsubsection{\textbf{Group Relative Policy Optimization}}
Group Relative Policy Optimization (GRPO)~\cite{deepseekmath} is a variant of Proximal Policy Optimization (PPO)~\cite{schulman2017proximalpolicyoptimizationalgorithms}.
Instead of relying on a learned value function, GRPO normalizes rewards at the group level, leveraging the mean and standard deviation of multiple sampled outputs for each prompt.
Recent studies have shown that GRPO achieves superior performance on verifiable tasks such as mathematical reasoning~\cite{r1}. 
Given that our relevance task also falls into the category of verifiable problems, GRPO is a natural and well-suited choice for our setting.
To optimize \(\pi_\theta\) with the objective Eq.~\eqref{eq:rl}, GRPO maximizes the following surrogate objective\footnote{Note that this is the on-policy version of the GRPO objective.}:
\begin{align}
\mathcal{J}_{\mathrm{GRPO}}(\theta) 
&= \mathbb{E}_{(p,q,n,l) \sim \mathcal{D}, \{\mathbf{o_i}\}_{i=1}^G \sim \pi_\theta(\cdot|p,q,n)} \nonumber \\& \Bigg[ \frac{1}{G} 
    \sum_{i=1}^G \frac{1}{|\mathbf{o}_i|} \sum_{t=1}^{|\mathbf{o}_i|} \log \pi_\theta(o_{i,t} \,|\, p,q,n, o_{i,<t}) \, \hat{A}_{i,t} \\& - \beta \;\mathbb{D}_\mathrm{KL}(\pi_{\theta} \| \pi_{\mathrm{ref}}) \nonumber\Bigg],
\end{align}
where \(\hat{A}_{i,t}\) is the advantage function estimated via group-based normalization:
\begin{equation}
\label{eq:adv}
\hat{A}_{i,t} = \frac{r_i - \mathrm{mean}(\mathbf{R})}{\mathrm{std}(\mathbf{R})},
\end{equation}
with \(\mathbf{R} = \{r_1, r_2, \ldots, r_G\}\) being the rewards of \(G\) sampled responses \(\{\mathbf{o}_1, \mathbf{o}_2, \ldots, \mathbf{o}_G\}\) for a given sample $(p,q,n)$.

\subsubsection{\textbf{Stepwise Advantage Masking}}
Despite GRPO's strong empirical performance, its group-wise normalization introduces a simplified form of sample-level advantage estimation, which in turn results in coarse-grained credit assignment.
In our task setting, such coarse supervision is particularly limiting: supervision solely at the trajectory level often fails to capture whether relevance criteria are correctly applied throughout the reasoning process, as errors made in intermediate steps cannot be accurately attributed or corrected.
However, traditional process supervision methods often require labor-intensive, full-trajectory annotations to train a process reward model~\cite{lightman2024lets}, or rely on computationally expensive Monte Carlo-based value estimation during RL training to provide additional online supervision signals~\cite{kazemnejad2025vineppo}.
To enable efficient process supervision in industrial-scale RL training, we heuristically instruct the model to provide intermediate scores at the end of each reasoning step using \verb|\boxed{}| via the prompt mentioned in Section~\ref{sec:prompt}.
This design allows us to extract process-level reward signals at no additional inference cost from a rule-based verifier.
Specifically, for each reasoning chain, we extract three relevance scores from the \verb|\boxed{}| outputs and use the positions of the boxed tokens as markers for different reasoning steps.
Specifically, for a query-note pair, let $s_i$ denote the predicted score at step $i \in {\{1, 2, 3\}}$\footnote{$s_3$ is exactly $\hat{l}$.}, and define $c_i = \texttt{True}$ if and only if $s_i = l$ (the ground-truth label).
For example, $(c_1,c_2,c_3)=(\texttt{False},\texttt{True},\texttt{True})$ means that the model failed to produce the correct result during general semantic analysis but succeeded after applying the specialized rules and maintained the correct result during the final reflection step.
This special prompt design enables us to examine not only the final prediction but also the correctness across reasoning steps.

Building on this setup, we propose a Stepwise Advantage Masking (SAM) mechanism (see Figure~\ref{fig:framework}).
Since the correctness of each reasoning step $(c_1, c_2, c_3)$ can be determined, we refine the application of the final advantage $A_t$, which is based on the outcome of $c_3$, by selectively attributing it only to the reasoning steps responsible for the final outcome.
Specifically:
(1) If the final CoT prediction is correct ($\hat{l} = l$), we reinforce only the correct reasoning steps (i.e., those with $c_i = \texttt{True}$), while ignoring incorrect ones to avoid reinforcing spurious reasoning;
(2) If the final CoT prediction is incorrect ($\hat{l} \neq l$), we penalize only the incorrect steps (i.e., those with $c_i = \texttt{False}$), masking out correct steps to avoid punishing valid reasoning segments.
Formally, we define a stepwise advantage mask $m_t$ for each token $t$ in the reasoning chain as:
\begin{equation}
m_t =
\begin{cases}
1, & \text{if } \hat{l} = l \ \land\ \text{step}(o_t) = i \ \land\ c_i = \texttt{True}, \\
1, & \text{if } \hat{l} \neq l \ \land\ \text{step}(o_t) = i \ \land\ c_i = \texttt{False}, \\
0, & \text{otherwise}.
\end{cases}
\label{eq:sam_mask}
\end{equation}
Here, $\text{step}(o_t)$ denotes the reasoning step index to which token $o_t$ belongs, determined by the boxed token positions.
The final policy objective of SAM-augmented GRPO is given by:
\begin{align}
\label{eq:sam_grpo}
\mathcal{J}^{SAM}_{\mathrm{GRPO}}(\theta) 
&= \mathbb{E}_{(p,q,n,l) \sim \mathcal{D}, \{\mathbf{o_i}\}_{i=1}^G \sim \pi_\theta(\cdot|p,q,n)} \nonumber \\& \Bigg[ \frac{1}{G} 
    \sum_{i=1}^G \frac{1}{|\mathbf{o}_i|} \sum_{t=1}^{|\mathbf{o}_i|}  \log \pi_\theta(o_{i,t} \,|\, p,q,n, \mathbf{o}_{i,<t}) \, \hat{A}_{i,t} \, \underline{\mathbf{m_{i,t}}} \\& - \beta \;\mathbb{D}_\mathrm{KL}(\pi_{\theta} \| \pi_{\mathrm{ref}}) \nonumber \Bigg],
\end{align}
where $\hat{A}_{i,t}$ denotes the advantage computed from the reward at the third step of the CoT path, and $\mathbf{m}_{i,t}$ is a binary mask that selectively modulates this advantage according to the correctness of $\text{step}(t)$.
We provide a pseudo-code describing our proposed SAM-augmented GRPO, as shown in Algorithm~\ref{alg:sam_grpo}.

\begin{algorithm}[t!]
\caption{SAM-augmented GRPO Training Procedure}
\label{alg:sam_grpo}
\KwIn{Training data $\mathcal{D} = \{(p, q, n, l)\}$, reference policy $\pi_{\mathrm{ref}}$, initial policy $\pi_{\theta}$, KL coefficient $\beta$, learning rate $\eta$.}
\KwOut{Updated policy model $\pi_{\theta}$.}

\While{training not converged}{
Sample a minibatch $\{(p, q, n, l)\} \sim \mathcal{D}$\;
\textbf{Experience Preparation:}\\
\For{each sample $(p,q,n,l)$}{
Generate $G$ responses $\{\mathbf{o}_i\}_{i=1}^{G} \sim \pi_{\theta}(\cdot|p,q,n)$\;
Compute rewards $\{r_i\}_{i=1}^G$ using Eq.~\eqref{eq:reward}\;
Estimate advantages $\{\hat{A}_i\}_{i=1}^G$ using Eq.~\eqref{eq:adv};

\For{each generated response $\mathbf{o}_i$}{
    Extract boxed scores $\{s_1, s_2, s_3\}$ and obtain correctness indicators $\{c_1, c_2, c_3\}$\;
    \For{each generated token $o_{i,t}$}{
    Identify associated step index $\text{step}(o_{i,t})$\;
    Compute advantage mask $m_{i,t}$ using Eq.~\eqref{eq:sam_mask}\;
    }
}
}
\textbf{Policy Update:}\\
Compute policy objective $\mathcal{J}^{SAM}_{\mathrm{GRPO}}(\theta)$ using Eq.~\eqref{eq:sam_grpo}\;
Update policy parameters $\theta \leftarrow \theta + \eta \nabla_{\theta}\mathcal{J}^{SAM}_{\mathrm{GRPO}}(\theta)$\;
}

\Return{$\pi_{\theta}$}\;
\end{algorithm}

\begin{table}
  \caption{Distribution of the two benchmark datasets.}
  \label{tab:datset}
  \setlength{\tabcolsep}{2mm}{
      \begin{tabular}{ccc}
        \toprule
        \textbf{Relevance Label} & \textbf{RANDOM}(\%) & \textbf{LONGTAIL}(\%)\\
        \midrule
        -1 & 2.54 & 2.75 \\
        0 & 37.40 & 45.69 \\
        1 & 6.60 & 7.28 \\
        2 & 19.03 & 15.71 \\
        3 & 34.43 & 28.57 \\
        \midrule
        total number & 15000 & 15000 \\
        \bottomrule
      \end{tabular}
      }
\end{table}

\section{Experiments}
\subsection{Experimental Settings}
\subsubsection{\textbf{Benchmarks and Metrics}}
Our experiments are conducted on two benchmark datasets widely used in real-world search deployments: \textsc{RANDOM} and \textsc{LONGTAIL}. 
Both datasets consist of 15,000 manually annotated query-note pairs, but differ in their query sampling distributions, enabling us to evaluate model robustness under both common and long-tail user behaviors.
The statistics of these datasets are summarized in Table~\ref{tab:datset}. 
We provide a detailed description of each dataset below:
\begin{itemize}
\item \textsc{RANDOM}: The queries are randomly sampled from a large-scale online search system, and for each query, notes are selected from candidate note lists at various ranking positions.
This design ensures that the dataset closely reflects the naturally occurring data distribution observed in practical deployments.
Each query-note pair is manually annotated and has undergone multiple rounds of quality inspection.
\item \textsc{LONGTAIL}: This dataset adopts the same configuration and sampling strategy for note selection as \textsc{RANDOM}, but differs in the query selection process: the queries are sampled based on their page view (PV) statistics, focusing on those that receive fewer than five PVs within a single day. These long-tail queries are inherently more challenging, demanding stronger semantic understanding for effective processing.
% \item \textbf{UNIFORM}: Also consisting of 15,000 query-note pairs, this dataset is constructed using the same sampling strategy as RANDOM. However, it differs by enforcing a strictly uniform class distribution, thereby eliminating class imbalance for evaluation.
\end{itemize}

We use accuracy (ACC), macro F1, and weighted F1 as our evaluation metrics. 
In particular, we report both 5-class ACC and 2-class ACC. 
The 2-class ACC is included to better align with our practical application needs, where labels $-1$ and $0$ are grouped into one category, and labels $1$, $2$, and $3$ are grouped into the other.
Reporting both metrics allows us to evaluate the model's fine-grained classification capability while also reflecting its practical decision-making behavior in real-world search applications.

\begin{table*}[!ht]
  \caption{
        Offline evaluation results.
  }
  \centering
  \label{tab:experiment_overall}
  \setlength{\tabcolsep}{1.65mm}{
      \begin{tabular}{cc|cccc|cccc}
                \toprule
                \multirow{2}{*}{\textbf{Model}}  & 
                \multirow{2}{*}{\textbf{Data}}  & 
                \multicolumn{4}{c|}{\textbf{\textsc{RANDOM}}} & \multicolumn{4}{c}{\textbf{\textsc{LONGTAIL}}} \\
                            
                &  & \textbf{2-ACC} &\textbf{5-ACC}
                &\textbf{Macro F1} &\textbf{Weighted F1} &\textbf{2-ACC} &\textbf{5-ACC} &\textbf{Macro F1} &\textbf{Weighted F1} \\
                \midrule
                    SFT-Label & 200k  & 90.26 & 78.64 & 65.66 & 77.40 & 89.11 & 77.66 & 63.24 & 76.31   \\   
                \midrule
                SFT-Reasoning-v1 & 150k & 80.10  & 59.44 & 47.52 & 59.65 & 77.50   & 59.22 & 47.45  & 60.34   \\
                SFT-Reasoning-v2 & 500k & 83.04 & 63.06  & 51.76 & 63.16 & 81.15  & 63.55  & 50.64 & 63.65  \\
                PPO-Reasoning & 150k+50k & 91.56 & 78.81 & 70.04 & 79.01 & 89.70 & 74.96 & 63.74 & 75.40 \\
                OutcomeRL-Reasoning & 150k+50k & 92.09 & 80.90 & 72.46 & 79.98 & 89.62 & 77.03 & 65.08 & 75.96 \\
                ProcessRL-Reasoning & 150k+50k & \textbf{92.45}  & \textbf{81.23}   &  \textbf{73.55}  & \textbf{80.43} & \textbf{90.04}   & \textbf{77.72} & \textbf{66.39} & \textbf{76.77} \\
                \bottomrule
        \end{tabular}
    }
\end{table*}

\subsubsection{\textbf{Data Preparation}}
\label{exp:trainingdata}

To construct the training dataset for our framework, we adopt the same query and note sampling strategy as used in the \textsc{RANDOM} benchmark. In total, we sampled 50K queries, each paired with 20 notes. 
25 professionally trained annotators labeled 1M query-note pairs over three months.
The overall label distribution is consistent with that of the \textsc{RANDOM} dataset.
To obtain the initial CoT reasoning data, we employ \texttt{DeepSeek-R1}~\cite{r1} to generate reasoning traces using the proposed criteria-augmented reasoning prompt.
Given that the raw reasoning outputs may be unreliable, we apply rejection sampling to retain only those traces whose final predictions are consistent with human annotations. 
Since the rejection sampling step inevitably alters the original label distribution, we slightly over-sample or down-sample certain instances to restore it to match the original dataset.
This process yields 500K validated reasoning trajectories.
To further enhance data quality in RL training, we introduce an offline difficulty estimation strategy based on \emph{avg@k}. For each query--note instance, we compute avg@k as the average prediction accuracy obtained by sampling the initial policy reasoning path $k$ times:
\begin{equation}
\text{avg@k} = \frac{1}{k}\sum_{i=1}^{k}\mathbb{I}(f(\mathbf{o}_i) = l).
\end{equation}
Intuitively, avg@k serves as a proxy for decision-boundary stability, reflecting the intrinsic difficulty of an instance.
We set $k=64$ in all experiments. Based on their avg@64 scores, we prune the training samples as follows:
\begin{itemize}
\item \emph{High-confidence samples}: those with $\text{avg@64} > 0.97$ (under the 5-class ACC setting) are discarded;
\item \emph{Low-confidence samples}: those with $\text{avg@64} < 0.04$ (under the 2-class ACC setting) are removed.
\end{itemize}
This asymmetric pruning strategy is not heuristic but essential: high-confidence samples contribute negligible gradient signals in RL optimization, while extremely low-confidence instances often originate from annotation errors or semantically insufficient notes that destabilize training. By removing both extremes, we retain a dataset that preserves the original distribution while emphasizing informative, yet non-trivial reasoning samples.

\subsubsection{\textbf{Models}}

Our base model, RedOne~\cite{zhao2025redonerevealingdomainspecificllm}, is an in-house post-trained version of \texttt{Qwen2.5-32B-Instruct}~\cite{qwen2025qwen25technicalreport}, further adapted to Xiaohongshu’s domain-specific search data.
To enable systematic comparison, we evaluate RedOne across different SFT and RL training methods, as well as across different scales of training data. 
% The evaluation focuses on determining whether our proposed process-supervised RL approach outperforms other baselines.

\begin{itemize}
    \item \textbf{SFT-Label}: We fine-tune RedOne on a sampled training set consisting of 200k (query, note, label) triples, where the model predicts the relevance label directly without generating intermediate reasoning traces.
    
    \item \textbf{SFT-Reasoning-v1}: We fine-tune RedOne on a sampled training set of 150k CoT data generated by \texttt{DeepSeek-R1}, where the model performs reasoning first and then outputs the relevance judgment.
    
    \item \textbf{SFT-Reasoning-v2}: This variant extends \textit{SFT-Reasoning-v1} with an additional 350k reasoning traces, totaling 500k samples, enabling training on a larger and more diverse reasoning corpus.

    \item \textbf{PPO-Reasoning}: We apply the standard PPO algorithm to optimize \textit{SFT-Reasoning-v1} using learned value heads for token-level advantage estimation. This setup enables more fine-grained optimization than outcome-based RL but may introduce potential bias in value estimation~\cite{yuan2025whatspposcollapselongcot}.

    \item \textbf{OutcomeRL-Reasoning}: We adopt the standard GRPO algorithm to optimize \textit{SFT-Reasoning-v1} in an outcome-based manner. The accuracy of the final relevance prediction serves as the reward for group computation, which results in uniform credit assignment and prevents differentiation across different steps of the reasoning process.
    
    \item \textbf{ProcessRL-Reasoning (ours)}: We apply SAM-augmented GRPO to optimize \textit{SFT-Reasoning-v1}, using partial advantage masking to enable step-level credit assignment and mitigate spurious reward propagation.
\end{itemize}

All RL-based methods are trained on the same 50k sampled training set, pruned using the avg@k-based offline difficulty estimation described in Section~\ref{exp:trainingdata}. 
The resulting dataset maintains a label distribution closely aligned with the \textsc{RANDOM} benchmark, ensuring fair and controlled comparisons across all RL variants.
Additional implementation details can be found in Appendix~\ref{apx:impl}.

\subsection{Offline Evaluation}
\subsubsection{\textbf{Overall Performance}}
As shown in Table~\ref{tab:experiment_overall}, our proposed ProcessRL-Reasoning approach achieves the best overall performance across both benchmarks and all evaluation metrics.

\textbf{\emph{RQ1: Does reasoning hurt or help with SFT?}}
We first compare the purely supervised baselines.  
The \textit{SFT-Label} (trained on 200k label-only instances) provides a strong reference point, reaching 90.26/78.64 (2-ACC/5-ACC) on \textsc{RANDOM} and 89.11/77.66 on \textsc{LONGTAIL}.  
In contrast, the two supervised reasoning variants, \textit{SFT-Reasoning-v1} (150k CoT traces) and \textit{SFT-Reasoning-v2} (500k CoT traces), perform notably worse: even with more data, \textit{SFT-Reasoning-v2} only attains 83.04/63.06 on \textsc{RANDOM} and 81.15/63.55 on \textsc{LONGTAIL}.  
This gap indicates that naively adding multi-step reasoning and training it with SFT alone does not automatically translate into better relevance modeling. Instead, it introduces additional optimization difficulty and error modes in the reasoning trajectory, which SFT alone cannot effectively correct.
This observation challenges the implicit assumption held by recent GRM studies—that incorporating multi-step reasoning naturally improves ranking quality—highlighting instead that reasoning without proper optimization can be detrimental, which motivates us to introduce RL-based optimization to better harness multi-step reasoning.

\textbf{\emph{RQ2: Can process-supervised RL truly enhance relevance reasoning?}}
We compare three RL variants grounded on the same SFT-initialized reasoning model. 
All variants use identical data, prompts, and reward definitions, ensuring that performance differences stem solely from how each method attributes credit during optimization.
\textit{PPO-Reasoning} delivers substantial improvements over \textit{SFT-Reasoning-v1}, but its reliance on value-function estimation introduces bias and unstable credit assignment, preventing it from fully exploiting the reasoning structure and leaving its performance close to the label-only baseline. 
In contrast, \textit{OutcomeRL-Reasoning} removes the value head entirely and normalizes rewards at the group level, thereby avoiding value estimation errors and stabilizing policy optimization. 
As a result, it further boosts performance to 80.90 5-ACC on \textsc{RANDOM} and 77.03 5-ACC on \textsc{LONGTAIL}, mostly outperforming PPO across all metrics.
However, outcome-based RL provides uniformly distributed advantages across all tokens within a trajectory, regardless of each step's contribution to the final decision.
Our \textit{ProcessRL-Reasoning} addresses this limitation by introducing stepwise credit assignment via the proposed SAM mechanism, mitigating spurious reward propagation. 
The \textit{ProcessRL-Reasoning} achieves the best overall performance—81.23 5-ACC and 73.55 Macro F1 on \textsc{RANDOM}, and 77.72 5-ACC and 66.39 Macro F1 on \textsc{LONGTAIL}—consistently surpassing both SFT and RL variants.
These results confirm that step-level credit assignment via SAM provides consistent gains over outcome-only RL, especially on Macro F1, demonstrating improved robustness and generalization in realistic search scenarios.

\subsubsection{\textbf{Analysis of data efficiency}}
\begin{figure}[!t]
    \centering
    \includegraphics[width=\linewidth]{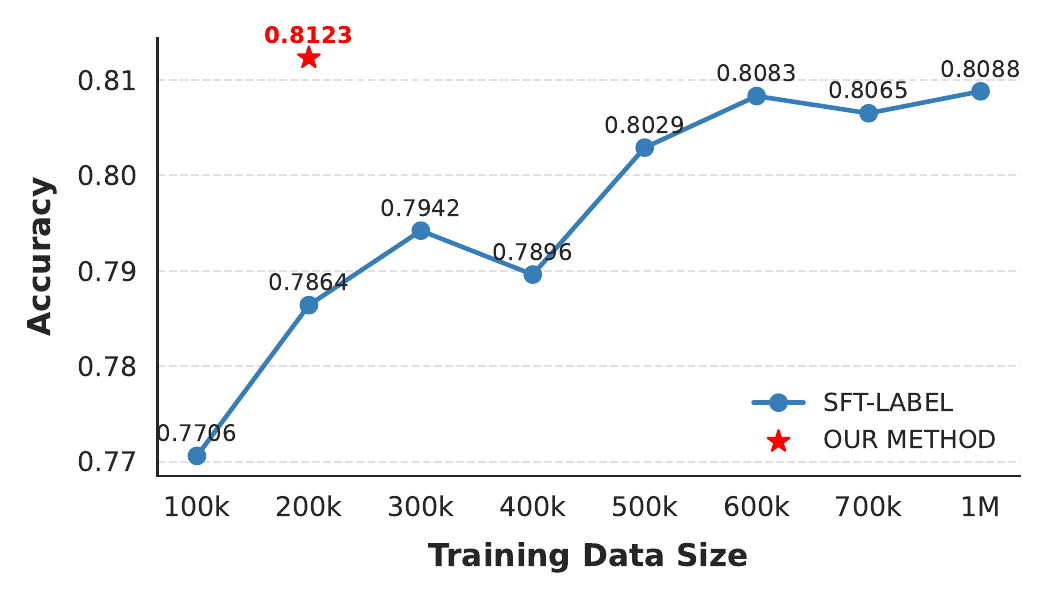}
    \caption{Analysis of data efficiency on SFT.}
    \label{fig:data_eff}
\end{figure}

To assess the training efficiency of our proposed approach, we conduct an ablation study comparing the performance of \textit{ProcessRL-Reasoning} against the \textit{SFT-Label} method using varying amounts of supervised training data. Specifically, we sample subsets of the labeled training data, ranging from 100K to 1M examples, and evaluate the resulting 5-class ACC of the SFT model on the \textsc{RANDOM} benchmark.
As shown in Figure~\ref{fig:data_eff}, the \textit{SFT-Label} baseline exhibits steady performance improvements as the training data size increases; however, its accuracy gains begin to plateau beyond 500K samples. 
In contrast, \textit{ProcessRL-Reasoning} achieves superior accuracy with far fewer samples. 
Remarkably, \textit{ProcessRL-Reasoning} outperforms the \textit{SFT-Label} model trained on 1M samples, using only 150K SFT CoT samples and 50K RL samples in total. 
This improvement can be attributed to the enhanced reasoning capabilities incentivized by process-supervised RL, which enables better generalization by leveraging stepwise reasoning rather than relying solely on rigid pattern matching, as seen in traditional SFT. 
In practical terms, our method requires far less labeled data to achieve competitive performance, highlighting its superior data efficiency. 
However, it still benefits from high-quality, accurately labeled training data to ensure effective optimization during the RL phase, underscoring the importance of robust data annotation and filtering strategies—such as our use of avg@k-based difficulty estimation.

\subsubsection{\textbf{Model dynamics between SFT and RL}}

As shown in Table~\ref{exp:ld}, \textit{ProcessRL-Reasoning} demonstrates superior F1-scores across most categories compared to \textit{SFT-label} with the same dataset size of 200k. In the 2-class setting, the improvement is particularly notable in the precision for the ``0*'' category (i.e., the aggregation of labels -1/0) and the recall for the ``1*'' category (i.e., the aggregation of labels 1/2/3), where the precision rate increases from 82.67 to 90.31, and the recall rate improves from 86.70 to 93.45.
This difference highlights that \textit{ProcessRL-Reasoning} is more cautious in identifying irrelevant examples and more proactive in recognizing potentially relevant ones, significantly addressing the \emph{under-recall} issue observed in \textit{SFT-label} when handling ambiguous positive examples.
From the perspective of the 5-class setting, the recall for the ``1*'' category improves through a general enhancement across labels 1/2/3, with only a slight sacrifice in precision for labels 2 and 3.
Furthermore, \textit{ProcessRL-Reasoning} shows significant improvements in both precision and recall for the minority classes (labels -1 and 1), demonstrating its strong ability to handle long-tail cases.
This improvement in both precision and recall indicates that our method refines the decision boundary by optimizing the reasoning process itself, rather than simply learning input-output mappings. 
These results suggest that incorporating grounded reasoning—where the model relies on domain-specific relevance criteria—enables the model to capture subtle distinctions between fine-grained categories, leading to improved classification performance.

\begin{table}[!h]
  \caption{The detailed differences between SFT and RL.}
  \label{exp:ld}
  \centering
  \label{tab:sftvsrl}
  \setlength{\tabcolsep}{1mm}{
      \begin{tabular}{c|ccc|ccc}
            \toprule
            \multirow{2}{*}{}  & \multicolumn{3}{c}{\textbf{\footnotesize SFT-label}} & \multicolumn{3}{c}{\textbf{\footnotesize ProcessRL-Reasoning}} \\
                        
           & \textbf{\footnotesize Precision} & \textbf{\footnotesize Recall} & \textbf{\footnotesize F1-Score} & \textbf{\footnotesize Precision} & \textbf{\footnotesize Recall} & \textbf{\footnotesize F1-Score} \\
            \midrule
            \textbf{\footnotesize 0*} & 82.67 & 94.43 & 88.16 & 90.31 & 90.96 & 90.63 \\
            \textbf{\footnotesize 1*} & 95.86 & 86.70 & 91.05 & 93.90 & 93.45 & 93.67 \\
            % \textbf{\footnotesize 2-ACC}   & - & -& 92.45 & - & - & 89.81 \\
            % \textbf{\footnotesize Macro F1}  &-  &- & 92.15 & - & - & 89.61 \\
            \midrule
            \textbf{\footnotesize -1} & 70.08 & 36.00 & 47.57 & 76.63 & 66.95 & 71.46 \\
            \textbf{\footnotesize 0}  & 78.87 & 94.33 & 85.91 & 87.89 & 89.52 & 88.69 \\
            \textbf{\footnotesize 1}  & 59.82 & 39.90 & 47.87 & 67.71 & 47.09 & 55.55 \\
            \textbf{\footnotesize 2} & 79.98 & 51.96 & 63.00 & 77.01 & 58.79 & 66.68 \\
            \textbf{\footnotesize 3} & 90.53 & 82.52 & 86.34 & 78.56 & 93.43 & 85.35 \\
            % \textbf{\footnotesize 5-ACC} & - & -& 81.23 & - & - & 79.24\\
            % \textbf{\footnotesize Macro F1}  &-  &- & 73.55 & - & - & 66.14 \\
            \bottomrule
    \end{tabular}
}
\end{table}

\subsection{Online Evaluation}
\subsubsection{\textbf{Online Deployment}}
To reduce inference latency in online deployment, we adopt a knowledge distillation strategy~\cite{44873}. 
Specifically, we use our RL-tuned model as a teacher to annotate millions of query--note pairs, and retain only the final-step relevance scores as supervision signals.
These signals are then distilled into a lightweight 0.1B-parameter BERT-based discriminative 5-class classifier, which serves as the student model for online inference.
In production, the deployed student model achieves a P95 latency of approximately 20ms with an almost 100\% response success rate.
% The predicted relevance scores are subsequently forwarded to the downstream ranking module, which determines the final ordering of notes in the Xiaohongshu app.
To preliminarily assess the effectiveness of the distilled student model, we conduct an offline evaluation on the \textsc{RANDOM} test set.
The student model attains a 2-ACC of 90.65, a 5-ACC of 79.22, a Macro F1 of 68.19, and a weighted F1 of 78.57.
These results consistently outperform the \textit{SFT-Label} baseline, while still exhibiting a noticeable gap compared to the teacher model.
This gap indicates the significant room for further improvement through more advanced distillation strategies.

% Notably, to further improve user experience in online systems, we adopt a binary classification (i.e., 2-class classification) setting, in which each note is labeled as either \textit{relevant} or \textit{irrelevant} according to its \textit{relevance score}. Relevant notes are prioritized in the ranking results, while irrelevant ones are relegated to lower positions.

\subsubsection{\textbf{Online Performance}}
To evaluate the student model online, we deployed it on the Xiaohongshu A/B testing platform. 
We randomly sampled 5\% of the online traffic as the test group and applied our proposed approach to this group. 
Another 5\% of the traffic was used as the baseline group, which continued to rely on the previous BERT-based model. 
For a fair comparison, we continuously monitored the experimental results on the platform. To mitigate the impact of traffic fluctuations, we set a minimum testing period of seven days.
Notably, we adopt the following two business metrics to evaluate online user experience:
\begin{itemize}
\item \textbf{CES}: CES is a core metric used to quantify user engagement on the platform. It aggregates user actions such as likes, favorites, comments, shares, and follows. A higher CES value indicates stronger user engagement and thus a better user experience.
\item \textbf{DCG 0/1}: DCG 0/1 is a key metric for evaluating ranking quality in search systems. In practice, a professional annotator manually reviews the top eight results from A/B experiments across 2,000 randomly sampled queries and labels each result as relevant or irrelevant. The metric counts the number of irrelevant results on the search result page, where a lower value indicates higher ranking quality and improved user experience.
\end{itemize}

According to the results in Table~\ref{table6}, our approach achieves a 0.72\% improvement in the CES metric, reflecting increased user engagement and an enhanced user experience. In addition, the DCG 0/1 metric shows a cumulative reduction of 0.36\%, indicating fewer irrelevant results and improved ranking quality. 
Overall, these findings demonstrate that our approach significantly enhances both user experience and ranking relevance.

\begin{table}[!htbp]
  \centering
  \caption{Online A/B test results (p < 0.05).}
  \setlength{\tabcolsep}{4mm}{
    \begin{tabular}{c|c|c}
      \toprule
      Model & $\Delta$CES $\uparrow$ & $\Delta$DCG 0/1 $\downarrow$ \\
      \midrule
      Baseline      & 0        & 0 \\
      Ours          & +0.72\%  & -0.36\% \\
      \bottomrule
    \end{tabular}
  }
\label{table6}
\end{table}

\section{Discussion}
\paragraph{\textbf{Over-Association and Reasoning Errors}}
We have identified \emph{overthinking} as a key failure mode, especially when handling ambiguous user queries about TV shows or movie characters. For example, if a query refers to Drama A but the note actually concerns Character C from Drama B (played by Actor D), the model might incorrectly associate the query with Drama B, since Actor D is also the lead in Drama A according to its internal knowledge.
This kind of spurious associative reasoning can sometimes lead to correct results in open-domain search, but it also risks generating false positives.
To address this, we plan to implement reasoning confidence modeling and introduce a refusal mechanism, aiming to minimize false inferences and improve the accuracy and stability of the reasoning process.

\paragraph{\textbf{Challenges in Criteria Adaptation}}
Our long-term goal is to train the model once and enable the business team to dynamically update criteria within the prompts, allowing the model to adapt to evolving business logic. 
We refer to this concept as \emph{Relevance LLM}. 
This model would be capable of adjusting to changing business requirements by leveraging a set of continuously updated rules.
However, our experiments show that the current RL-tuned model is overfitted to the fixed rule set used during training. 
When the rules are modified during inference, the model still tends to reason based on the original logic learned during training. 
We hypothesize this happens because the criteria system was static during RL training, without exposure to dynamic rule changes. 
Future work will focus on introducing dynamic criteria variations during training to improve robustness and reduce overfitting, ensuring the model can handle criteria modifications more effectively during inference.

\paragraph{\textbf{Combining SAM with LLM-as-Verifier}}
In this work, we propose SAM, an intuitive process-supervised strategy that enhances outcome-based RL by enabling step-level credit assignment. The success of SAM relies on unbiased stepwise correctness judgments, achieved through exact matching in our scenario. However, such rule-based verification may not be feasible for more general reasoning tasks. With the rise of LLMs as generative verifiers~\cite{DBLP:conf/iclr/ZhangHBKKA25}, LLMs offer a powerful, domain-agnostic approach to verifying each reasoning step in a more flexible and generalizable manner. Combining SAM with LLMs as verifiers could present a promising direction for future research, allowing for more robust verification and credit assignment in general RL tasks. 
\citet{xie2025capoenhancingllmreasoning} has already conducted preliminary studies in this area.

\section{Conclusion}
In this paper, we introduce a reinforcement learning paradigm that formulates relevance modeling in Xiaohongshu search as a multi-step reasoning task. 
By integrating domain-specific relevance criteria into structured prompts and proposing Stepwise Advantage Masking (SAM) for step-level credit assignment, 
our method delivers grounded and interpretable relevance reasoning, especially in complex and ambiguous search scenarios.
Extensive offline evaluations and online A/B tests demonstrate that our approach achieves significant improvements across key metrics, providing a practical solution for industrial relevance modeling.

% \begin{acks}
% \end{acks}

\bibliographystyle{ACM-Reference-Format}
\bibliography{sample-base}

\appendix

\section{Effectiveness of Relevance Criteria}
\label{apx:effect_rel_cri}
To validate the effectiveness of our multi-step reasoning prompt with integrated domain-specific relevance criteria, we perform zero-shot reasoning with \texttt{DeepSeek-R1} on the \textsc{RANDOM} benchmark.
As shown in Table~\ref{exp:RC}, incorporating the relevance criteria into the prompt substantially improves performance across all metrics, including 2-ACC, 5-ACC, and Macro F1, compared with the setting without relevance criteria.
These improvements highlight the importance of domain-specific rules in enhancing the accuracy and reliability of GRMs in real-world search scenarios.

\begin{table}[!h]
  \caption{
       Impact of relevance criteria on zero-shot reasoning.
  }
  \centering
  \label{exp:RC}
    \begin{tabular}{c|ccc}
        \toprule
        \multirow{2}{*}{\textbf{Prompt}}  & 
        \multicolumn{3}{c}{\textbf{\textsc{RANDOM}}}\\
        & \textbf{2-ACC} & \textbf{5-ACC} & \textbf{Macro F1}  \\
        \midrule
        w/o relevance criteria & 68.70  & 41.50 & 37.10 \\
        w/ relevance criteria  & 71.20  & 49.30 & 40.23 \\
        \bottomrule
    \end{tabular}
\end{table}

\section{Implementation Details}
\label{apx:impl}
Our training pipeline consists of two stages: supervised fine-tuning and reinforcement learning.

\paragraph{Supervised Fine-Tuning}
We fine-tune each model for one epoch using 16 Nvidia H800 GPUs with a global batch size of 64. 
Training is performed using the AdamW optimizer with a fixed learning rate of \(2 \times 10^{-6}\). 
The maximum sequence length is set to 8192 tokens.

\paragraph{Reinforcement Learning}
We adopt the VERL framework~\cite{10.1145/3689031.3696075} and perform training on 40 Nvidia H800 GPUs, using a batch size of 40 and a mini-batch size of 40 to enable on-policy optimization.
To ensure sufficient coverage of all label categories in the early phase of RL training, we curate a 30k subset from a 50k RL dataset, aiming for an approximately uniform label distribution.
We train the policy model on this subset for 3 epochs with a learning rate of \(1 \times 10^{-6}\) and 20 warm-up steps.
In the later phase, we switch to the original 50k dataset, which reflects the real-world online distribution. 
We reduce the learning rate to \(8 \times 10^{-7}\) and continue training for 1 epoch with a warm-up ratio of 0.03. 
For GRPO, the group size is set to 8. 
For PPO, the critic learning rate is set to \(1 \times 10^{-6}\) with 20 warm-up steps. 
All learning rates follow a cosine decay schedule.
The KL coefficient for PPO and GRPO is set to \(1 \times 10^{-3}\). 
The maximum input and output lengths during RL training are 7168 and 2048 tokens, respectively. 
Rollout sampling uses a temperature of 1.0 and a top-p of 0.95. 
Validation is conducted on 10\% of the \textsc{RANDOM} dataset, and the best checkpoints are retained throughout the training process.

% \subsection{}

\end{document}